\documentclass{article}
\begin{document}

\title{ON THE QUARK-QUARK SUPERCONDUCTING PAIRING INDUCED BY INSTANTONS IN QCD}
\author{N. V. Hieu, H. N. Long, N. H. Son and N. A. Viet\\
{\small Institute of Physics, P. O. Box 429, Boho, Hanoi 10000, Vietnam.}}
\date{}
\maketitle
\begin{abstract}
The superconducting pairing of quarks induced by instantons in QCD is
studied by means of the functional integral method. The integral equation
determining the critical temperature of the superconducting phase transition
is established. It is shown that the Bose condensate of diquarks consists of
color antitriplet $\left( \bar{3}\right) $ scalar diquarks.
\end{abstract}

The superconducting pairing of quarks in QCD was investigated
since more than twenty years$^{[1-5]}$. At the present time a strong
interest to this problem is motivated by recent works of Alford, Rajagopal
and Wilczeck$^{\left[ 6\right] }$, Sch\"{a}fer and Wilczek$^{\left[ 7\right]
}$, Rapp, Sch\"{a}fer, Shuryak and Velkovsky$^{\left[ 8\right] }$ and other
authors$^{\left[ 9-16\right] }$. In this work the functional integral method$%
^{\left[ 15\right] }$ is applied to the study of the superconducting pairing
of quarks induced by instantons$^{\left[ 8,17,18\right] }$, and the integral
equation determining the critical temperature of phase transition is established.\par
We work in the imaginary time formalism and use the notations 
\[
x=\left( \mathbf{x},\tau \right) ,\qquad \int dx=\int_{0}^{\beta }d\tau \int
d\mathbf{x},\qquad \beta =\frac{1}{kT}, 
\]
$k$ and $T$ being the Boltzmann constant and the temperature. Denote $\psi
_{A}\left( x\right) =\psi _{A}\left( \mathbf{x},\tau \right) $ the quark
field, $A$ being the set consisting of the index $\alpha $ of the Dirac
spinor, the color and flavor indices $a$ and $i$, $A=\left( \alpha ai\right) 
$, $a=1,2,...\,,\,N_{c}$, $i=1,2,\,...\,,\,N_{f}$. Consider the system of
massless quarks (and antiquarks) with the parity conserving and chirally
symmetric interaction Lagrangian of the form 
\begin{eqnarray}
L_{int} &=&\frac{1}{2}\overline{\psi }^{A}\left( x\right) \overline{\psi }%
^{B}\left( x\right) U_{BA}^{CD}\psi _{D}\left( x\right) \psi _{C}\left(
x\right) ,  \label{1} \\
U_{BA}^{CD} &=&-U_{AB}^{CD}=-U_{BA}^{DC}=U_{AB}^{DC}.  \label{2}
\end{eqnarray}
The partition function of the system equals 
\begin{eqnarray}
Z &=&\int \left[ D\psi \right] \left[ D\overline{\psi }\right] \exp \left\{
-\int dx\overline{\psi }^{A}\left( x\right) D_{A}^{B}\psi _{B}\left(
x\right) \right\}  \nonumber \\
&&\exp \left\{ \frac{1}{2}\int dx\,\overline{\psi }^{A}\left( x\right) 
\overline{\psi }^{B}\left( x\right) U_{BA}^{CD}\psi _{D}\left( x\right) \psi
_{C}\left( x\right) \right\}  \label{3}
\end{eqnarray}
with 
\begin{equation}
D_{A}^{B}=\left[ \gamma _{4}\left( \frac{\partial }{\partial \tau }-\mu
\right) +\mathbf{\gamma \nabla }\right] _{\alpha }^{\beta }\delta
_{a}^{b}\delta _{i}^{j},  \label{4}
\end{equation}
$\mu $ being the chemical potential. Introducing the antisymmetric bi-spinor
fields $\Phi _{DC}\left( x\right) ,\,\overline{\Phi }^{AB}\left( x\right) ,$%
\begin{equation}
\Phi _{CD}\left( x\right) =-\Phi _{DC}\left( x\right) ,\,\,\overline{\Phi }%
^{BA}\left( x\right) =-\overline{\Phi }^{AB}\left( x\right) ,  \label{5}
\end{equation}
and the functional integral over these fields 
\begin{equation}
Z_{0}^{\Phi }=\int \left[ D\Phi \right] \left[ D\overline{\Phi }\right] \exp
\left\{ -\frac{1}{2}\int \overline{\Phi }^{AB}\left( x\right)
U_{BA}^{CD}\Phi _{DC}\left( x\right) \right\} ,  \label{6}
\end{equation}
we have the Hubbard-Stratonovich transformation$^{\left[ 12, 15\right] }$%
\begin{eqnarray}
&&\exp \left\{ \frac{1}{2}\int dx\,\overline{\psi }^{A}\left( x\right) 
\overline{\psi }^{B}\left( x\right) U_{BA}^{CD}\psi _{D}\left( x\right) \psi
_{C}\left( x\right) \right\}  \nonumber \\
&&\hspace{2cm}=\frac{1}{Z_{0}^{\Phi }}\int \left[ D\Phi \right] \left[ D%
\overline{\Phi }\right] \exp \left\{ -\frac{1}{2}\int \overline{\Phi }%
^{AB}\left( x\right) U_{BA}^{CD}\Phi _{DC}\left( x\right) \right\}  \label{7}
\\
&&\hspace{2.5cm}\exp \left\{ -\frac{1}{2}\int dx\left[ \overline{\Delta }%
^{CD}\left( x\right) \psi _{D}\left( x\right) \psi _{C}\left( x\right) +%
\overline{\psi }^{A}\left( x\right) \overline{\psi }^{B}\left( x\right)
\Delta _{BA}\left( x\right) \right] \right\} ,  \nonumber
\end{eqnarray}
where 
\begin{equation}
\Delta _{BA}\left( x\right) =U_{BA}^{CD}\Phi _{DC}\left( x\right) ,\,\,%
\overline{\Delta }^{CD}\left( x\right) =\overline{\Phi }^{AB}\left( x\right)
U_{BA}^{CD}.  \label{8}
\end{equation}
Applying this transformation in the r.h.s. of the relation (3) and
performing the functional integration over fermionic variables, we rewrite
the partition function $Z$ in the form of a functional integral over the
bosonic diquark fields $\Phi _{DC}\left( x\right) $ and $\overline{\Phi }%
^{AB}\left( x\right) $ : 
\begin{equation}
Z=\frac{Z_{0}}{Z_{0}^{\Phi }}\int \left[ D\Phi \right] \left[ D\overline{%
\Phi }\right] \exp \left\{ I_{eff}\left[ \Phi ,\overline{\Phi }\right]
\right\} ,  \label{9}
\end{equation}
$Z_{0}$ being the partition function of the system of the free quarks, 
\begin{equation}
Z_{0}=\int \left[ D\psi \right] \left[ D\overline{\psi }\right] \exp \left\{
-\int dx\overline{\psi }^{A}\left( x\right) D_{A}^{B}\psi _{B}\left(
x\right) \right\} .  \label{10}
\end{equation}
The effective action of the diquark system equals 
\begin{eqnarray}
I_{eff}\left[ \Phi ,\overline{\Phi }\right] &=&-\frac{1}{2}\int \overline{%
\Phi }^{AB}\left( x\right) U_{BA}^{CD}\Phi _{DC}\left( x\right) +W\left[
\Delta ,\overline{\Delta }\right] ,  \label{11} \\
W\left[ \Delta ,\overline{\Delta }\right] &=&\sum_{n=1}^{\infty }W^{\left(
2n\right) }\left[ \Delta ,\overline{\Delta }\right] ,  \label{12}
\end{eqnarray}
\begin{equation}
W^{\left( 2\right) }\left[ \Delta ,\overline{\Delta }\right] =\frac{1}{2}%
\int dx_{1}\int dx_{2}\overline{\Delta }^{D_{1}C_{1}}\left( x_{1}\right)
S_{C_{1}}^{B_{2}}\left( x_{1}-x_{2}\right) \Delta _{B_{2}A_{2}}\left(
x_{2}\right) S_{D_{1}}^{T\,A_{2}}\left( x_{2}-x_{1}\right) ,  \label{13}
\end{equation}
\begin{eqnarray}
W^{\left( 4\right) }\left[ \Delta ,\overline{\Delta }\right] &=&-\frac{1}{4}%
\int dx_{1}\int dx_{2}\int dx_{3}\int dx_{4}\overline{\Delta }%
^{D_{1}C_{1}}\left( x_{1}\right) S_{C_{1}}^{B_{2}}\left( x_{1}-x_{2}\right)
\Delta _{B_{2}A_{2}}\left( x_{2}\right)  \nonumber \\
&&S_{D_{3}}^{T\,A_{2}}\left( x_{2}-x_{3}\right) \overline{\Delta }%
^{D_{3}C_{3}}\left( x_{3}\right) S_{C_{5}}^{B_{4}}\left( x_{3}-x_{4}\right)
\Delta _{B_{4}A_{4}}\left( x_{4}\right) S_{D_{1}}^{T\,A_{4}}\left(
x_{4}-x_{1}\right) ,\hspace{1cm}  \label{14} \\
........ &&  \nonumber
\end{eqnarray}
where $S_{A}^{B}\left( x-y\right) $ is the two-point Green function of the
free quark, 
\begin{equation}
D_{A}^{B}S_{B}^{C}\left( x-y\right) =\delta _{A}^{C}\left( x-y\right) ,
\label{15}
\end{equation}
and 
\begin{equation}
S_{B}^{TA}\left( x-y\right) =S_{B}^{A}\left( y-x\right) .  \label{16}
\end{equation}

From the variational principle 
\begin{equation}
\frac{\delta I_{eff}\left[ \Phi ,\overline{\Phi }\right] }{\delta \overline{%
\Phi }^{AB}\left( x\right) }=0  \label{17}
\end{equation}
we derive the system of field equations 
\begin{eqnarray}
\Delta _{A_{1}B_{1}}\left( x_{1}\right)
&=&U_{A_{1}B_{1}}^{D_{1}C_{1}}\left\{ \int dx_{2}S_{C_{1}}^{A_{2}}\left(
x_{1}-x_{2}\right) \Delta _{A_{2}B_{2}}\left( x_{2}\right)
S_{D_{1}}^{T\,B_{2}}\left( x_{2}-x_{1}\right) \right.  \nonumber \\
&&-\int dx_{2}\int dx_{3}\int dx_{4}S_{C_{1}}^{A_{2}}\left(
x_{1}-x_{2}\right) \Delta _{A_{2}B_{2}}\left( x_{2}\right)
S_{D_{3}}^{T\,B_{2}}\left( x_{2}-x_{3}\right)  \label{18} \\
&&\qquad \left. \overline{\Delta }^{D_{3}C_{3}}\left( x_{3}\right)
S_{C_{3}}^{A_{4}}\left( x_{3}-x_{4}\right) \Delta _{A_{4}B_{4}}\left(
x_{4}\right) S_{D_{1}}^{T\,B_{4}}\left( x_{4}-x_{1}\right) +...\right\}
.\qquad  \nonumber
\end{eqnarray}
The constant solutions 
\begin{eqnarray*}
\Delta _{AB}\left( x\right) &=&\Delta _{AB}^{0}=\mathrm{const}, \\
\overline{\Delta }^{DC}\left( x\right) &=&\overline{\Delta }^{0DC}=\mathrm{const}
\end{eqnarray*}
of this system of equations are the order parameters of the superconducting
phase transition with the formation of the Bose condensate of diquarks - the
Cooper pairs of quarks. At the critical temperature $T\rightarrow T_{c}$, $%
\beta \rightarrow \beta _{c}$, these order parameters tend to vanishing
limits and in the r.h.s. of the equations (18) we can neglect the high order
terms. Then we obtain the system of linear equations 
\begin{equation}
\Delta _{AB}^{0}=U_{AB}^{DC}\int dxS_{C}^{A^{\prime }}\left( -x\right)
\Delta _{A^{\prime }B^{\prime }}^{0}S_{D}^{TB^{\prime }}\left( -x\right) .
\label{19}
\end{equation}
Using the expression of the two - point Green function of the free quarks at 
$\beta \rightarrow \beta _{c}$%
\begin{equation}
S_{A}^{B}\left( x\right) =\frac{1}{\beta _{c}}\sum_{m}\exp \left\{
i\varepsilon _{m}\tau \right\} \frac{1}{\left( 2\pi \right) ^{3}}\int d%
\mathbf{p}\exp \left\{ i\mathbf{px}\right\} \frac{\left( -i\right) \left[
\left( \varepsilon _{m}+i\mu \right) \gamma _{4}+\mathbf{p\gamma }\right]
_{\alpha \beta }}{\left( \varepsilon _{m}+i\mu \right) ^{2}+\mathbf{p}^{2}},
\label{20}
\end{equation}
\[
\varepsilon _{m}=\left( 2m+1\right) \frac{\pi }{\beta _{c}}, 
\]
$m$ being the integers, and introducing the $4\times 4$ matrices $\widehat{%
\Delta }_{\left( ai\right) \left( bj\right) }^{0}$ with the elements 
\[
\left[ \widehat{\Delta }_{\left( ai\right) \left( bj\right) }^{0}\right]
_{\alpha \beta }=\Delta _{\left( \alpha ai\right) \left( \beta bj\right)
}^{0}, 
\]
we rewrite the equations (19) in the matrix notations

\begin{eqnarray}
\left[ \widehat{\Delta }_{\left( ai\right) \left( bj\right) }^{0}\right]
_{\alpha \beta } &=&U_{\left( \alpha ai\right) \left( \beta bj\right)
}^{\left( \delta d\ell \right) \left( \gamma ck\right) }\frac{1}{\beta _{c}}%
\sum_{m}\frac{1}{\left( 2\pi \right) ^{3}}\int d\mathbf{p}  \nonumber \\
&&\frac{\left[ \left\{ \left( \varepsilon _{m}+i\mu \right) \gamma _{4}+%
\mathbf{p\gamma }\right\} \widehat{\Delta }_{\left( ak\right) \left( d\ell
\right) }^{0}\left\{ \left( \varepsilon _{m}-i\mu \right) \gamma _{4}+%
\mathbf{p\gamma }\right\} ^{T}\right] _{\gamma \delta }}{\left[ \left(
\varepsilon _{m}+i\mu \right) ^{2}+\mathbf{p}^{2}\right] \left[ \left(
\varepsilon _{m}-i\mu \right) ^{2}+\mathbf{p}^{2}\right] }.\hspace{1cm}
\label{21}
\end{eqnarray}
The order parameter matrix with the elements $\Delta _{\left( \alpha
ai\right) \left( \beta bj\right) }^{0}$ and the coupling constant matrix
with the elements $U_{\left( \alpha ai\right) \left( \beta bj\right)
}^{\left( \delta d\ell \right) \left( \gamma ck\right) }$ have the general
forms$^{\left[ 15\right] }$%
\begin{equation}
\Delta _{\left( \alpha ai\right) \left( \beta bj\right) }^{0}=\Delta
_{\left( ai\right) \left( bj\right) }^{S}\left( \gamma _{5}C\right) _{\alpha
\beta }+\Delta _{\left( ai\right) \left( bj\right) }^{P}\left( C\right)
_{\alpha \beta },  \label{22}
\end{equation}
\begin{eqnarray}
U_{\left( \alpha ai\right) \left( \beta bj\right) }^{\left( \delta d\ell
\right) \left( \gamma ck\right) } &=&\frac{1}{4}S_{\left( ai\right) \left(
bj\right) }^{\left( d\ell \right) \left( ck\right) }\left( \gamma
_{5}C\right) _{\alpha \beta }\left( C^{-1}\gamma _{5}\right) ^{\delta \gamma
}+\frac{1}{4}P_{\left( ai\right) \left( bj\right) }^{\left( d\ell \right)
\left( ck\right) }\left( C\right) _{\alpha \beta }\left( C^{-1}\right)
^{\delta \gamma }  \nonumber \\
&&+\frac{1}{16}V_{\left( ai\right) \left( bj\right) }^{\left( d\ell \right)
\left( ck\right) }\left( \gamma _{\mu }\gamma _{5}C\right) _{\alpha \beta
}\left( C^{-1}\gamma _{5}\gamma _{\mu }\right) ^{\delta \gamma }+\frac{1}{16}%
A_{\left( ai\right) \left( bj\right) }^{\left( d\ell \right) \left(
ck\right) }\left( \gamma _{\mu }C\right) _{\alpha \beta }\left( C^{-1}\gamma
_{\mu }\right) ^{\delta \gamma }\hspace{1cm}  \nonumber \\
&&+\frac{1}{24}T_{\left( ai\right) \left( bj\right) }^{\left( d\ell \right)
\left( ck\right) }\left( \sigma _{\mu \nu }\gamma _{5}C\right) _{\alpha
\beta }\left( C^{-1}\gamma _{5}\sigma _{\mu \nu }\right) ^{\delta \gamma }.
\label{23}
\end{eqnarray}
Substituting these expressions into the r.h.s of the equations (21), we
derive two equations for $\Delta _{\left( ai\right) \left( bj\right) }^{S}$
and $\Delta _{\left( ai\right) \left( bj\right) }^{P}$ : 
\label{24}
$$
\Delta _{\left( ai\right) \left( bj\right) }^{S} =-JS_{\left( ai\right)
\left( bj\right) }^{\left( d\ell \right) \left( ck\right) }\Delta _{\left(
ck\right) \left( d\ell \right) }^{S},  \eqno (24.a)$$
$$
\Delta _{\left( ai\right) \left( bj\right) }^{P} =JP_{\left( ai\right)
\left( bj\right) }^{\left( d\ell \right) \left( ck\right) }\Delta _{\left(
ck\right) \left( d\ell \right) }^{P},  \eqno (24.b)$$
\setcounter{equation}{24}
where 
\begin{equation}
J=\frac{1}{\beta _{c}}\sum_{m}\frac{1}{\left( 2\pi \right) ^{3}}\int d%
\mathbf{p}\frac{\varepsilon _{m}^{2}+\mathbf{p}^{2}+\mu ^{2}}{\left[ \left(
\varepsilon _{m}+i\mu \right) ^{2}+\mathbf{p}^{2}\right] \left[ \left(
\varepsilon _{m}-i\mu \right) ^{2}+\mathbf{p}^{2}\right] }.  \label{25}
\end{equation}
It can be shown that 
\begin{equation}
J=\frac{1}{8\pi ^{2}}\int_{0}^{\lambda }\left[ \frac{1}{p-\mu }\tanh \frac{%
\beta _{c}\left( p-\mu \right) }{2}+\frac{1}{p+\mu }\tanh \frac{\beta
_{c}\left( p+\mu \right) }{2}\right] p^{2}dp.  \label{26}
\end{equation}
The ultraviolet divergence in the integral (25) may be avoided by
introducing the momentum cut-off $\lambda $.

In the case of the effective four-quark direct coupling induced by instantons%
$^{\left[ 8,17,18\right] }$ we have 
\begin{eqnarray}
&&U_{\left( \beta bj\right) \left( \alpha ai\right) }^{\left( \gamma
ck\right) \left( \delta d\ell \right) }=\frac{1}{2}\delta _{a}^{c}\delta
_{b}^{d}\left[ \delta _{i}^{k}\delta _{j}^{\ell }-\left( \mathbf{\tau }%
\right) _{i}^{k}\left( \mathbf{\tau }\right) _{j}^{\ell }\right] \left\{
G_{1}\left[ \delta _{\alpha }^{\gamma }\delta _{\beta }^{\delta }+\left(
\gamma _{5}\right) _{\alpha }^{\gamma }\left( \gamma _{5}\right) _{\beta
}^{\delta }\right] +G_{2}\left( \sigma _{\mu \nu }\right) _{\alpha }^{\gamma
}\left( \sigma _{\mu \nu }\right) _{\beta }^{\delta }\right\}
\hspace*{0.75cm}  \nonumber \\
&&\hspace*{1.5cm}-\frac{1}{2}\delta _{a}^{d}\delta _{b}^{c}\left[ \delta
_{i}^{\ell }\delta _{j}^{k}-\left( \mathbf{\tau }\right) _{i}^{\ell }\left( 
\mathbf{\tau }\right) _{j}^{k}\right] \left\{ G_{1}\left[ \delta _{\alpha
}^{\delta }\delta _{\beta }^{\gamma }+\left( \gamma _{5}\right) _{\alpha
}^{\delta }\left( \gamma _{5}\right) _{\beta }^{\gamma }\right] +G_{2}\left(
\sigma _{\mu \nu }\right) _{\alpha }^{\delta }\left( \sigma _{\mu \nu
}\right) _{\beta }^{\gamma }\right\}  \label{27}
\end{eqnarray}
with 
\begin{equation}
G_{1}=g\frac{1}{N_{c}^{2}-1}\frac{2N_{c}-1}{2N_{c}},\,\,\,G_{2}=-g\frac{1}{%
N_{c}^{2}-1}\frac{1}{4N_{c}},  \label{28}
\end{equation}
where $g$ is some positive constant. It follows that 
\begin{equation}
S_{\left( ai\right) \left( bj\right) }^{\left( d\ell \right) \left(
ck\right) }=P_{\left( ai\right) \left( bj\right) }^{\left( d\ell \right)
\left( ck\right) }=\frac{1}{4}G\varepsilon _{ij}\varepsilon ^{k\ell }\left(
\delta _{a}^{c}\delta _{b}^{d}-\delta _{a}^{d}\delta _{b}^{c}\right)
\label{29}
\end{equation}
with 
\begin{equation}
G=g.\frac{1}{2N_{c}\left( N_{c}-1\right) }.  \label{30}
\end{equation}
Substituting the expression (29) of $S_{\left( ai\right) \left( bj\right)
}^{\left( d\ell \right) \left( ck\right) }$ and $P_{\left( ai\right) \left(
bj\right) }^{\left( d\ell \right) \left( ck\right) }$ into the r.h.s. of the
equations (24.a) and (24.b), we derive the formulae 
\begin{eqnarray}
\Delta _{\left( ai\right) \left( bj\right) }^{S} &=&\varepsilon _{ij}\Delta
_{\left[ ab\right] }^{S},  \nonumber \\
\Delta _{\left( ai\right) \left( bj\right) }^{P} &=&\varepsilon _{ij}\Delta
_{\left[ ab\right] }^{P},  \label{31}
\end{eqnarray}
where $\Delta _{\left[ ab\right] }^{S}$ and $\Delta _{\left[ ab\right] }^{P}$
are antisymmetric under the permutation of the color indices and satisfy the
equations 
\label{32}
$$
\Delta _{\left[ ab\right] }^{S}\left( 1-GJ\right) =0,  \eqno (32.a)$$
$$
\Delta _{\left[ ab\right] }^{P}\left( 1+GJ\right) =0.  \eqno (32.b)$$
\setcounter{equation}{32}
\vskip -0.1truecm \noindent%
Because both constants $G$ and $J$ are positive, $\Delta _{\left[ ab\right]
}^{P}$ must be zero 
\begin{equation}
\Delta _{\left[ ab\right] }^{P}=0.  \label{33}
\end{equation}
\vskip -0.1truecm \noindent%
For the existence of a non-vanishing antisymmetric rank 2 spinor $\Delta
_{\left[ ab\right] }^{S}$ the equation 
\begin{equation}
GJ=1  \label{34}
\end{equation}
\vskip -0.1truecm \noindent%
must be satisfied. In another form this equation was derived by Pisarski and
Rischke$^{\left[ 19\right] }$. In some approximation it was discussed in ref$%
^{\left[ 8\right] }$. \noindent Together with the expression (26) for $J$
the equation (34) determines the critical temperature $T_{c}$: 
\begin{equation}
T_{c}\sim \mathrm{const }\lambda \exp \left\{ -\frac{8\pi ^{2}}{\mu ^{2}G}%
\right\} .  \label{35}
\end{equation}
\vskip -0.1truecm \noindent%
The existence of $\Delta _{\left[ ab\right] }^{S}\neq 0$ is a consequence of
the attractive interaction between two quarks inside a scalar antitriplet $%
\left( \overline{\mathrm{3}}\right) $ diquark. The relation (33) means that
the pseudoscalar diquark does not exist.%
\vskip 0.2cm
\noindent {\bf ACKNOWLEDGMENT}
\vskip 0.1cm
The authors express their sincere thank to the Natural Science
Council for the financial support to this work.


\end{document}